\documentclass[a4paper, 11pt]{paper}
\usepackage{enumerate}   
\usepackage{graphicx}
\usepackage{hyperref}
\usepackage{authblk}

\hypersetup{
    bookmarks=true,         
    unicode=false,          
    pdftoolbar=true,        
    pdfmenubar=true,        
    pdffitwindow=false,     
    pdfstartview={FitH},    
    pdftitle={Who is Who in Phylogenetic Networks: Articles, Authors and Programs},    
    pdfauthor={Tushar Agarwal, Philippe Gambette, David Morrison},     
    pdfnewwindow=true,      
    colorlinks=true,       
    linkcolor=red,          
    citecolor=green,        
    filecolor=magenta,      
    urlcolor=cyan           
}
\begin{document}

\title{Who is Who in Phylogenetic Networks: Articles, Authors and Programs}

\author{\normalsize Tushar \textsc{Agarwal}$^{1}$, Philippe \textsc{Gambette}$^{2,\ast}$, David \textsc{Morrison}$^{3}$}

\affil[]{\footnotesize{$^1$Computer Science and Engineering, Indian Institute of Technology Ropar, Punjab 140001, India;}\\
\footnotesize{$^{2}$Universit{\'e} Paris-Est, LIGM (UMR 8049), UPEM, CNRS, ESIEE, ENPC, Marne-la-Vall{\'e}e 77454, France;}\\
\footnotesize{$^{3}$Systematic Biology, EBC, Uppsala University, Norbyv\"{a}gen 18D, Uppsala 75236, Sweden} \\
\footnotesize{$^{\ast}$Correspondence to be sent to Philippe Gambette (\href{mailto:philippe.gambette@u-pem.fr}{philippe.gambette@u-pem.fr})}\\
}
\date{\today} 

\maketitle

\abstract{The phylogenetic network emerged in the 1990s as a new model to represent the evolution of species in the case where coexisting species transfer genetic information through hybridization, recombination, lateral gene transfer, etc. As is true for many rapidly evolving fields, there is considerable fragmentation and diversity in methodologies, standards and vocabulary in phylogenetic network research, thus creating the need for an integrated database of articles, authors, techniques, keywords and software. We describe such a database, ``Who is Who in Phylogenetic Networks'', available at \url{http://phylnet.univ-mlv.fr}. ``Who is Who in Phylogenetic Networks'' comprises more than $600$ publications and $500$ authors interlinked with a rich set of more than $200$ keywords related to phylogenetic networks. The database is integrated with web-based tools to visualize authorship and collaboration networks and analyze these networks using common graph and social network metrics such as centrality (betweenness, eigenvector, degree and closeness) and clustering. We provide downloads of raw information about entries in the database, and a facility to suggest modifications and contribute new information to the database. We also present in this article common use cases of the database and identify trends in the research on phylogenetic networks using the information in the database and textual analysis.

\textbf{Keywords:} phylogenetic network, co-authorship network, social networks, phylogenetic networks software.
}

\normalsize
\section*{Introduction}
The phylogenetic network has gained popularity in the past two decades as a model of evolution capable of accounting for vertical inheritance as well as gene flow events such as recombination, hybridization and horizontal gene transfer, which the simple tree-based model fails to do. The rapid evolution of the body of knowledge related to, and the community working on, phylogenetic networks has led to significant variety in many phylogenetic methodologies, such as reconstruction and classification, and software to compute, evaluate, compare, and visualize phylogenetic networks. Although some surveys~\cite{Lapointe2000,MKL2006,Morrison2010b,HusonScornavacca2011,Nakhleh2013,Morrison2014a} and books~\cite{HRS2011,Morrison2011,DHKSM2012,Gusfield2014} describe this variety, ``Who is Who in Phylogenetic Networks'' is the first resource of its kind aiming to consolidate this diversity into an easily navigable database. 

The website was started in 2007 as an interactive bibliographic database  of phylogenetic networks research, based on an open source project, BibAdmin (\url{https://gforge.inria.fr/projects/bibadmin/}). The code has been extensively extended  to incorporate capabilities to draw word clouds and network visualizations. New graphs and sections were also added to the website to present the raw information in the database in a more meaningful way. 

In addition to providing information about the community working on phylogenetic networks, ``Who is Who in Phylogenetic Networks'' makes available an easy-to-use interface to study the evolution of a new, swiftly growing scientific community. This database and the associated website are intended to serve as an encyclopedic and bibliographic resource of information about the community, tools, models and methods related to phylogenetic networks. Such a resource has inherent value for current members of the phylogenetic community, particularly researchers and students new to the field, as well as the community working on biological methods in general.

\section*{Authors and Keywords in the Database}

\subsection*{Authors}

Each author's entry in the database has associated with it a picture, the author's country of origin as determined automatically from the domain name of the author's website, a link to the author's personal or professional homepage, as well as a link to view his or her publications in Google Scholar. Each publication's entry in the database consists of, in addition to essential data about the publication itself, a set of keywords, a DOI link, and an abstract, if available.

\subsection*{Keywords}

Publications in the database are manually tagged by keywords, which help identify articles or methods based on:
(i) the nature of the problem solved in an article (reconstruction, consensus, labeling, comparison, generation, visualization, etc.) and its computational complexity (NP-complete, APX-hard, polynomial); 
(ii) the nature of algorithms given in an article (``exponential algorithm'' and ``FPT'' -- for ``fixed-parameter tractable'' -- are approaches to solving NP-hard problems exactly, whereas ``approximation'' or ``heuristic'' algorithms aim to find approximate solutions); 
(iii) the nature of input data; 
(iv) the restrictions on subclasses of networks studied in the article (e.g. ``galled trees'', ``tree-child networks'', etc.); 
(v) the existence of an implementation, indicated by the tag ``software''; 
(vi) the names of software, articles and methods used or implemented (keywords starting with the word ``Program'').
A definition is provided for technical keywords. Furthermore, keywords act as a convenient and meaningful navigational aid in the website, as described in several use cases below.

\section*{Use cases}

The main page of the ``Who is Who in Phylogenetic Networks'' highlights four use cases, corresponding to an action users might wish to take or a type of information they might wish to obtain: find experts, explore research, discover software and follow the community. 




\subsection*{Find Experts}

This section comprises two types of co-authorship network visualizations: first, a set of precomputed graphs augmented with various social network metrics; and second, a set of dynamic graphs with metrics computed directly on the user's computer. 

Each node in a precomputed graph is tagged with measures of degree, authority, hub, number of triangles containing that node as a vertex, betweenness centrality, closeness centrality, eigenvector centrality, clustering coefficient and eccentricity (for definitions, see e.g.~\cite{Estrada2015}). All of these metrics are computed using the open source network analysis tool Gephi~\cite{Bastian2009}. The web interface makes it possible for the user to see the global evolution of the phylogenetic network community by displaying the network corresponding to any year from 1990 through 2015. 

\begin{figure}[!h]
\begin{center}
\includegraphics[scale=0.22]{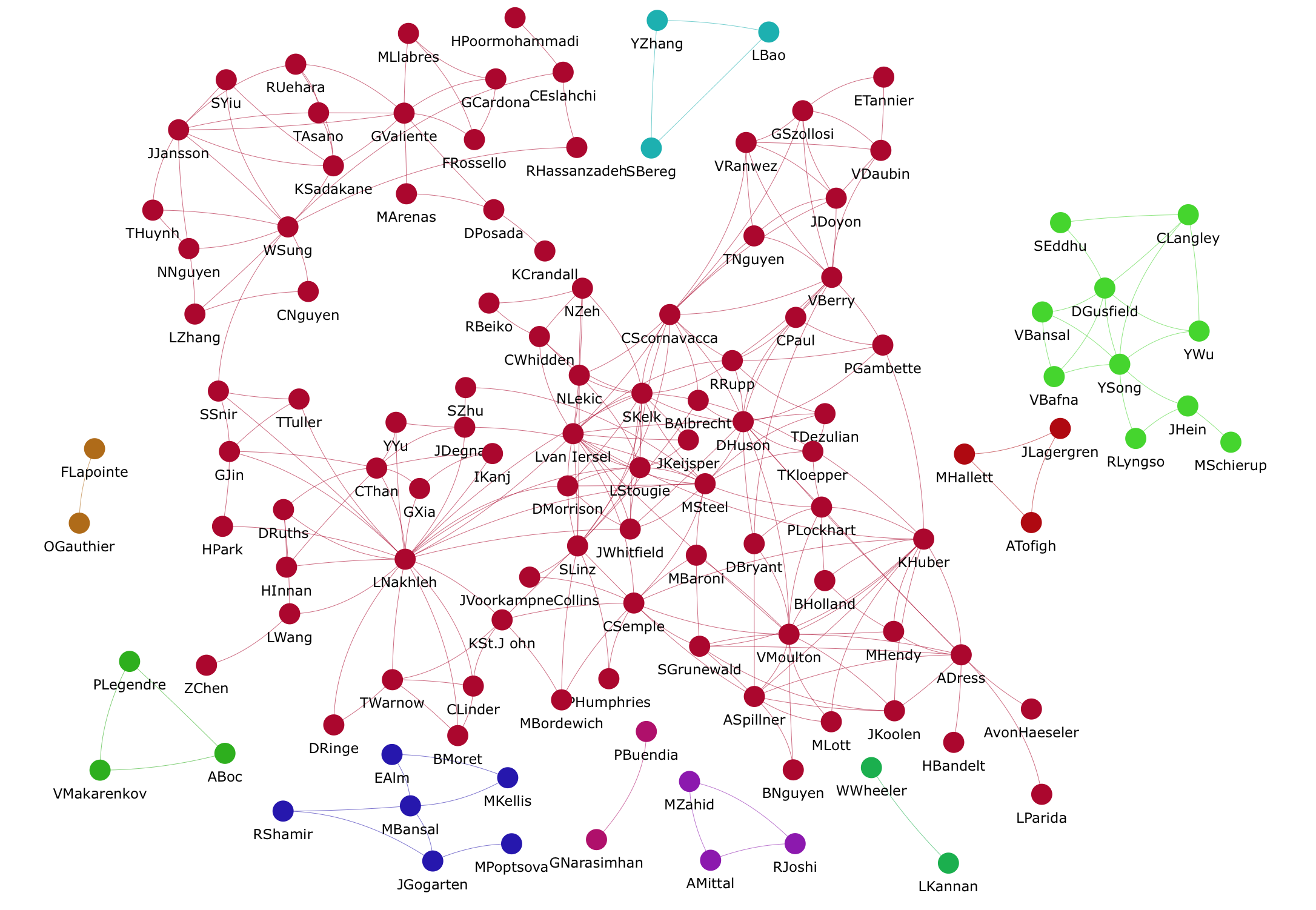}
\end{center}
\caption{\label{fig:dynamicGraphs} A screenshot of a portion of the 2014 co-authorship network in ``Who is Who in Phylogenetic Networks''.}
\end{figure}

The dynamic graphs (Fig.~\ref{fig:dynamicGraphs}) provide the user sufficient flexibility to display only those nodes fulfilling customizable structural and temporal constraints. It is possible to display the co-authorship network for any arbitrary range of time, and focus only on authors with a configurable minimum number of publications. Nodes can be colored along a linear gradient depending on the values
of the degree, the number of triangles containing that node as a vertex, or the centrality measures mentioned above. This makes the dynamic graph amenable to visual analysis, and enables easy identification of significant nodes in the community. For instance, the social network metrics available enable visual identification of nodes that are ``central'' in their communities, linked to many other nodes, or act as  ``bridges'' between communities.  Nodes that do not pass the specified criterion of a chosen minimum number of publications are ignored in the calculation of the color gradient. The user may also choose to highlight the authors belonging to a particular country, or see a detailed research profile and the list of indexed articles of an author, simply by clicking on the corresponding node.

Information about journals and conferences the phylogenetic networks community publishes in or meets at is presented in the form of tag clouds.
In addition to providing an exhaustive listing of publication venues, these give visual hints about the relative importance of journals and conferences, and may help to identify possible venues for submission, or find articles on phylogenetic methods. The size and color of a word in a tag cloud varies across a logarithmic scale, with larger size and warmer colors representing a greater number of publications in the corresponding journal or conference.


The entire community working on phylogenetic networks is depicted pictorially as a collage of authors' photos. The dimensions of an author's picture represent the number of publications by that author weighted by the number of co-authors on each publication.

The ``Find Experts'' section also consists of a co-authorship network visualization enriched with information about keywords in the database, and with the capability to color nodes, representing authors, across a gradient depending on their ``focus'', the fraction of total publications of the corresponding author tagged with a keyword, or their ``prolificacy'', the total number of publications of the corresponding author tagged with a keyword, if they satisfy a customizable threshold of minimum number of publications.

\subsection*{Explore Research}

Publications in the database are accompanied by information on their media of publication or presentation, such as books, journals, conferences and theses. This section includes bar charts illustrating the trend of all indexed publications about phylogenetic networks or those including a particular keyword.

The ``Explore research'' section also consists of a word-cloud of keywords, and also a ``tree cloud''~\cite{Gambette2010} plus a natural extension of this, a ``network cloud'' built using NeighborNet~\cite{Bryant2004neighbor}. These provide a global view of the database, and display all keywords structured in the form of a tree depicting their semantic relationships, automatically deduced from their co-occurrence in the database. Here, too, the size and color of a word in a word-cloud varies along a logarithmic scale, with larger size and warmer color representing a greater number of publications containing the corresponding keyword.

\subsection*{Discover Software}

The structure of the website and the presence of keywords makes it possible for biologists to identify one or more programs suitable to study their data, which might be in a particular format or structure. A network depicting relationships between input data and programs aids this process. Each keyword corresponding to a certain type of input data (starting with ``from'', such as ``from sequences'', ``from distances'', ``from rooted trees'', etc.) links to keywords representing software (starting with ``Program'', such as ``Program SplitsTree'', ``Program Phylonet'', ``Program TCS'', etc.). Each program has a short description including a download link.


\subsection*{Follow Community}

Using the ``Follow community'' feature, a user can receive an update whenever a particular author's article is indexed or when a new article is published by any author in the community.

\section*{Mapping the phylogenetic networks research and community}

In this section, we demonstrate how the ``Who is Who in Phylogenetic Networks'' database can be used to
analyze trends about the community of researchers working on phylogenetic networks. We focus first on their collaborations, and second on their research interests.

\subsection*{Evolution of the Co-authorship Network}

The co-authorship network provided on the ``Who is Who in Phylogenetic Networks'' website shows how individual research groups working on phylogenetic networks started developing inter-group research collaborations, which are represented as connected components in the graph. Increasing collaboration results in the merging of connected components over the years from 1991 through 2015. In fact, it is natural that, as more and more publications are taken into account to build this network, its largest connected component (a ``giant'' component) becomes larger with time.

In order to analyze in greater detail the evolution of the global structure of the co-authorship network, we focus on uniform-sized time slices, rather than consider all publications added since the oldest one. Therefore, we focus on the 2005-2015 period, where the number of publications is great enough (more than 30 articles per year). We generated the co-author graph for slices or periods of 5 years starting from 2005 (2005 through 2009, 2006 through 2010 and so on), connecting two authors if they have published at least three publications together in a slice or period. We chose this threshold because it implies that the linked researchers have worked on more than one project together; as the conference versions of journal papers are referenced on the website, choosing a threshold of two publications would not have been high enough for this purpose.

The main connected component of the graph reveals interesting information about the structure of collaborations over the years\footnote{The corresponding complete dataset is available at \url{https://goo.gl/vA8U6G}.}.
For instance, in the 2005-2009, 2006-2010 and 2007-2011 periods, it always contains only Luay Nakhleh and his co-authors. Furthermore, Luay Nakhleh is an author of almost all publications (except one) explaining the edges in these connected components. However, for the later years, the main connected components do not contain such a single vertex that is adjacent to all of the other vertices. Furthermore, the size of the connected component in the first three periods is at most ten, whereas for the three last periods, it ranges between 12 and 15. This shows that in recent years, several distinct groups, especially in Europe (France, the Netherlands, the UK, Germany and Sweden), have started collaborating to conduct research on phylogenetic networks.

\subsection*{Mapping the Phylogenetic Network Research Topics}

We also study the trends and characteristics of phylogenetic networks research topics, by using the Digital Object Identifiers (DOIs) stored in the ``Who is Who in Phylogenetic Networks'' database. The DOIs allow easy access to the abstracts of the publications, using scientific databases such as the Web of Science (\url{https://www.webofknowledge.com/}) or Scopus (\url{https://www.scopus.com/}). Again, we focus on the 2005-2015 period, to have a high enough and stable number of publications in each year.
Using Scopus, we found the abstracts for 305 publications in this period.

\begin{figure}[!h]
\begin{center}
\includegraphics[scale=0.36]{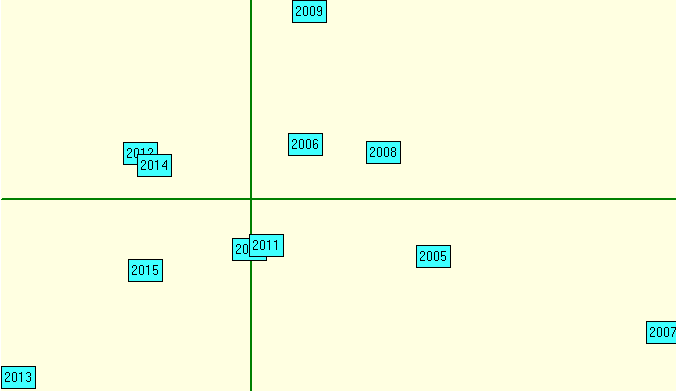}
\end{center}
   \caption{\label{fig:yearAFC} Factor analysis of the abstracts of
   305 publications about phylogenetic networks in the 2005-2015 period,
   grouped by year.}
\end{figure}

First, we performed a factor analysis on the abstracts grouped together by year (see Fig.~\ref{fig:yearAFC}), using Lexico 3 (\url{http://www.lexi-co.com}). In the factor graph, we find a time trend, with the years 2005 through 2009 on the right side, years 2010 and 2011 in the middle, and years 2012 through 2015 on the left side. Vocabulary that is significantly overrepresented in the 2005-2009 period includes ``recombination'', ``hgt'', ``sequences'', ``consensus'' and ``metrics''. In the 2012-2015 period, ``reconciliation'', ``trinets'', ``cost'', ``duplication'', ``loss'' and ``binary'' are overrepresented. This reflects the recent development of efficient methods to reconstruct reconciliation scenarios taking into account not only lateral gene transfer but also duplication and loss, as well as the introduction of new combinatorial objects related to networks, like trinets.

\begin{figure}[!h]
\begin{center}
\includegraphics[scale=0.36]{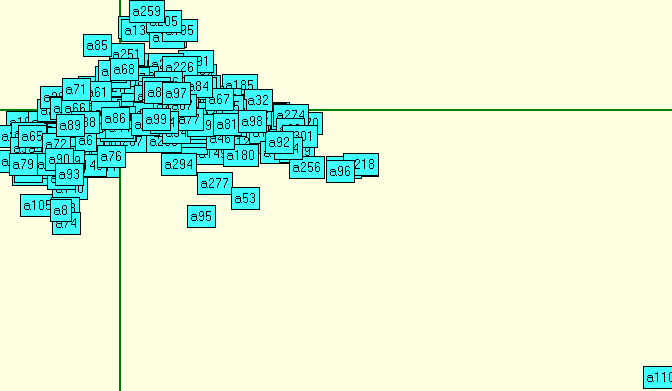}
\end{center}
   \caption{\label{fig:docAFC} Factor analysis of the abstracts of
   305 publications about phylogenetic networks in the 2005-2015 period.}
\end{figure}

Second, we performed the factor analysis of all documents, as shown in Fig.~\ref{fig:docAFC}. The graph highlights interesting characteristics of the phylogenetic networks research topics. The main axis distinguishes between the more algorithmic or mathematical papers on the right side (``n'', ``networks'', ``algorithm'', ``phylogenetic'', ``rooted'', ``time'', ``vertices'', ``number'') and the more biological papers on the left side (``gene'', ``hgt'', ``reconciliation'', ``transfer'', ``methods'', ``inference'', ``species'', ``lineage'').
The second axis seems to distinguish between approaches that explicitly reconstruct abstract or explicit phylogenetic networks at the top (``networks'', ``split'', ``phylogenetic'', ``circular'', ``'class', ``taxa'', ``cluster'', ``trinets'', ``quartet''), and reconciliation scenarios, with a special focus on the model, at the bottom (``gene'', ``hgt'', ``reconciliation'', ``lineage'', ``lgt'', ``transfer'', ``species'', ``duplication'', ``model'').
Note that the isolated paper in the bottom right corner is very mathematical and deals with proving a formula, with this formula presented in this article's abstract, and this explains its unusual position in the representation. Our results do not change if we remove this article's abstract from the corpus.

A similar kind of analysis can be used on a subset of publications to focus on a more specific topic, and this can be useful, for example, when composing a research article, where it may be helpful to provide an overview of the research done so far, or the state of the art.

\section*{Availability}

``Who is Who in Phylogenetic Networks'' is freely available online at \url{http://phylnet.univ-mlv.fr}. The database and website code is open source and available at \url{https://github.com/tushar-agarwal/phylnet}.

\section*{Conclusion}

We described ``Who is Who in Phylogenetic Networks'', a database associated with web-based tools that allow users to explore the community of researchers working on phylogenetic networks and the research done in this field. Not only is the database of inherent value in trying to present the diversity in the methodologies and vocabulary in research on phylogenetic networks in a single location, but it is useful both for researchers working on designing new methods involving phylogenetic networks, and for bioinformaticians who wish to obtain an overview of existing methods to perform analyses on their data. We also showed how the content present in this database can be used for general analyses about the research, and the social network of researchers, in this field. 

\bibliography{main}
\bibliographystyle{alpha}

\end{document}